%Paper: astro-ph/9408081
%From: karsten@ulysses.llnl.gov (Karsten Jedamzik)
%Date: Tue, 23 Aug 94 17:57:20 PDT

\lineskip=3pt minus 2pt
\lineskiplimit=3pt
\magnification=1200
\centerline{\bf BARYON NUMBER TRANSPORT IN A COSMIC QCD-PHASE
TRANSITION}
\vskip 0.7in
\centerline{{Karsten Jedamzik$^{a,b,\dag}$} and
{George M. Fuller$^{b}$}}
\vskip 0.12in
\centerline{\sl ${\rm {}^a}$ Physics Research Program}
\centerline{\sl Institute for Geophysics and Planetary Physics}
\centerline{\sl University of California}
\centerline{\sl Lawrence Livermore National Laboratory}
\centerline{\sl Livermore, CA 94550}
\vskip 0.12in
\centerline{\sl ${\rm {}^b}$ Department of Physics}
\centerline{\sl University of California, San Diego}
\centerline{\sl La Jolla, CA 92093-0319}
\vskip 1.3in
\centerline{\bf ABSTRACT}
\baselineskip=12pt plus 1pt

We investigate the transport of baryon number across
phase boundaries
in a putative first order cosmic QCD-phase transition.
Two independent phenomenological models are employed
to estimate the baryon penetrability at the
phase boundary: chromoelectric
flux tube models; and an analogy to baryon-baryon coalescence
in nuclear physics.
Our analysis indicates that baryon transport across phase boundaries
may be order of magnitude more efficient than other work has
suggested. We discuss the substantial uncertainties involved in
estimating baryon penetrability at phase boundaries.

\vskip 1.7in
\centerline {\sl ${}^{\dag}$ Present address: University of California,
Lawrence Livermore National Laboratory, L413,}
\centerline {\sl Livermore, CA 94550}

\vfil\eject
\baselineskip=18pt plus 2pt

\centerline{\bf 1. Introduction}
\vskip 0.15in

The possibility of a cosmic first order QCD-phase transition
[1,2]
has generated considerable interest.
Such a phase transition would occur when the universe cools to a temperature
of order $T\sim 100$MeV. During a {\it first order} QCD phase transition
we would expect macroscopic separation of phases, so that macroscopic
bubbles of \lq\lq quark-gluon plasma\rq\rq\ phase coexist in rough
pressure equilibrium with a \lq\lq hadron\rq\rq\ phase of distinct
color-singlet particles.

It has been speculated that a first order QCD-phase transition in the
early universe could produce observable
remnants. In particular, possible formation
of baryon number inhomogeneities during such a first order
transition has been considered extensively
[1,2,3,4,5].
Baryon number inhomogeneities which persist to the epoch $T\sim 100$keV
could affect primordial nucleosynthesis yields. Primordial nuclear
abundances in such inhomogeneous models can be very different from those
in a homogeneous standard Big Bang.
There has been extensive research on the effects of baryon
inhomogeneities on the epoch of primordial nucleosynthesis [3,4,5,6].
Observationally inferred primordial abundances may allow us to constrain
aspects of inhomogeneity production in cosmic phase transitions.

Inhomogeneities could also lead to the formation of stable or
metastable strange-quark nuggets
[1,7].
It has been suggested that stable nuggets could be a natural candidate
for dark matter. The existence
of nuggets could also have implications for the epoch of primordial
nucleosynthesis [8].
It has even been suggested that
primordial black holes [9]
and primordial magnetic fields [10] could originate in
a first order QCD-phase transition.

Remnants from a cosmic QCD-phase
transition, such as baryon inhomogeneities, strange quark matter
nuggets, primordial black holes, and primordial magnetic fields,
could only be produced if the transition is {\it first order}.
Recent QCD lattice calculations seem to favor a higher
order transition [11]. However, the numerical
calculations are still restricted to rather small lattices and
thus do not correctly model the continuum limit.
Another
complication in lattice models comes from uncertainty in the
numerical treatment of light
quarks [12]. A definitive answer on the order of the
QCD-transition probably will come only with the development of a future
generation of computers.

If the QCD-transition is of first order, there will
be an associated macroscopic separation of phases.
Crudely, we can describe each component as either a
\lq\lq deconfined\rq\rq\ quark-gluon plasma phase, or a
\lq\lq confined\rq\rq\ hadron phase.
During the phase transition these components will be able to coexist
in thermodynamic and chemical equilibrium at a coexistence
temperature, $T_c$. Bonometto and Pantano [13] have given a good review of
many of the phase transition issues.

In the early universe an idealized scenario for a first order
QCD-phase transition might be as follows [5].
Initially the universe is
at high temperature and in the quark-gluon plasma phase.
The net baryon number reside entirely in the quark-gluon
plasma and is distributed homogeneously. Eventually, the cosmic
expansion will cool the universe to temperature $T_c$ and
small bubbles of
a hadronic phase appear for the first time. Subsequently, the cosmic
expansion will require a continuous conversion of quark-gluon plasma
to hadron phase. The rate of conversion between quark-gluon plasma
and hadron phase is determined by the requirement that the universe
be kept at (or very near) the coexistence temperature, $T_c$. The phase
transition is
completed when all the quark-gluon plasma has been converted to
hadron phase. At this point, all the baryon number
resides in the hadron phase and is distributed homogeneously.
This scenario assumes a universe
which is in complete thermal and chemical equilibrium
during the transition.

In reality, a cosmic first order QCD-phase transition will necessarily
result in
deviations from thermal and chemical equilibrium. The
magnitude of these deviations will depend on the efficiency of heat
and baryon transport during the phase transition.

There are two mechanism for heat transport: hydrodynamic flow and
neutrino heat conduction [2].
In the case of hydrodynamic flow, the
quark-gluon plasma phase will maintain a slightly higher pressure
than the hadronic phase. This pressure gradient will cause
a fluid flow, and an associated energy flow, from the
quark-gluon plasma to the hadron phase.
Quarks and antiquarks in the quark-gluon plasma will convert into
color singlet mesons on the phase boundary.
This conversion of quarks and antiquarks into mesons will involve details
of strong
interaction physics.

In the case of neutrino heat conduction, the quark-gluon plasma
will maintain a slightly higher temperature than the hadronic phase.
The temperature gradient will cause a heat flow from the quark-gluon
plasma to the hadron phase. Heat will be transported most efficiently
by neutrinos. This is because neutrinos have relatively long mean free paths.
Heat transport by neutrinos does not involve strong interaction
physics at the phase boundary, in contrast to the case of heat
transport by hydrodynamic flow.
Which heat transport mechanism is the dominant one depends on
the equation of state of the phases, the physical properties of
the phase boundary, and the geometry of the phase boundary
[14].
Heat transport by neutrinos could be favored over heat
transport by hydrodynamic flow if there is an inhibition in the
formation of mesons at the phase boundary.

The transport of baryon number from the quark-gluon plasma into
the hadron phase necessarily involves strong interaction processes
at the phase boundary.
This is because
weakly interacting neutrinos can not carry baryon number.

In the present paper we estimate the baryon number penetrability
at the phase boundary. This baryon number penetrability we denote as
$\Sigma_h$. Specifically, $\Sigma_h$ is the probability that a
baryon
which approaches the phase boundary from the hadronic phase,
such as a proton, neutron, or baryonic resonance,
will dissociate into quarks and pass over into the quark-gluon
plasma. Alternatively, we could estimate the probability that a quark
which approaches the phase boundary from the quark-gluon plasma
will form a color singlet baryon at the phase boundary. We denote this
probability
as $\Sigma_q$. The thermal averages of the probabilities,
$\langle\Sigma_h\rangle$ and $\langle\Sigma_q\rangle$, are related
by detailed balance [5]
$$\kappa f_{q-\bar q}\langle\Sigma_q\rangle = f_{b-\bar b}
\langle\Sigma_h\rangle\ .\eqno(1)$$
Here $f_{q-\bar q}$ is the excess quark flux, i.e. the flux of
quarks minus the flux of antiquarks.
Analogously,
$f_{b-\bar b}$ is the excess baryon flux.

The dimensionless
quantity $\kappa$ will take values between $\kappa=1/3$ and
$\kappa=1$. In the limit where $\kappa=1/3$, baryon number predominantly
passes over the phase boundary as pre-formed three-quark clusters in
the quark-gluon phase.
We would expect this limit to obtain for quark-gluon plasmas
which are highly correlated.
The value $\kappa=1/3$ then accounts for
the fractional baryon number ($1/3$) of a single quark. In the limit
where $\kappa =1$, baryons will be predominantly formed by single
energetic quarks which cross over the phase
boundary into the hadron phase.
We will estimate the efficiency of this process
in detail in Section 2.

The maximum possible value of $\langle\Sigma_h\rangle$ is unity,
i.e. all baryons which approach
the phase boundary from the hadron phase
dissociate into three free quarks and pass
into the quark-gluon plasma.
A value of $\langle\Sigma_h\rangle =1$ is possible when there
is no energy threshold associated with this process.
However, the quarks in the
baryon have to \lq\lq rearrange\rq\rq\ themselves into quasi-free quarks
at the phase boundary. This rearrangment process might involve an
energy barrier. Thus it is conceivable that $\langle\Sigma_h\rangle$
is well below unity.

It should be mentioned that baryon number inhomogeneities can form
during a first order QCD-phase transition even for maximum baryon
number penetrability, $\langle\Sigma_h\rangle =1$.
When bubbles of quark-gluon plasma are small in size (i.e., towards
the end of the phase transition) phase boundaries will move at high
velocities. If phase boundaries move rapidly compared to baryon
number transport rates, then the chemical equilibrium in baryon number
across the phase boundaries may break down. In essence, a shrinking bubble of
quark-gluon plasma may shrink so fast that the baryon number inside of it
is \lq\lq trapped\rq\rq\ and turned into hadrons in situ.

In this way baryons could be concentrated in the
shrinking quark-gluon plasma bubbles. Note that this process starts
from chemical equilibrium. Even in equilibrium,
baryon number
preferentially resides in the quark-gluon phase [4].
Net baryon number densities in the quark-gluon plasma can exceed
net baryon number densities in the hadron phase by up to several orders
of magnitude, even if both phases are in chemical
equilibrium with each other. Thus, towards the end of the phase transition a
large
fraction of the net baryon number may still reside in
shrinking bubbles of quark-gluon plasma.

Between the middle and the end of the phase transition the typical
phase boundary velocity will increase continuously with time. The velocity
for which baryon number transport is too slow to establish chemical
equilibrium near the phase boundary is determined by bulk
properties of the phases and the baryon number penetrability,
$\langle\Sigma_h\rangle$. A low value for $\langle\Sigma_h\rangle$
implies an early drop-out of chemical equilibrium and thus might
lead to the formation of very high-amplitude baryon number
inhomogeneities.
Such a scenario might lead to the formation of strange quark matter
nuggets or other remnants.

In Section 2 we will estimate $\langle\Sigma_h\rangle$
with the help of phenomenological chromoelectric flux tube models.
Such a
calculation has been done previously by Sumiyoshi {\it et al.} [15].
We will find results quite different than those obtained by Sumiyoshi {\it et
al.}.
In Section 3 we employ an analogy to baryon-baryon
coalescence in nuclear physics
in order to give an independent estimate of $\langle\Sigma_h\rangle$.
We give conclusions in Section 4.
\vskip 0.15in
\centerline{\bf 2. Chromoelectric Flux Tube Models}
\vskip 0.1in

Chromoelectric flux tube models provide us with a phenomenological
understanding for the formation of hadrons from initially
\lq\lq free\rq\rq\ quarks.
These models assume the existence of a chromoelectric field between
any two oppositely colored quarks. The chromoelectric field strength is assumed
to be constant in magnitude and independent of the separation of
the quarks. These fields can be thought of as being confined to a tube of
constant width.
This is referred to as flux tube. Results of lattice QCD justify
the assumption of such a flux tube [16].
Chromoelectric flux tube models
are successful in explaining observations of processes such as
$e^+e^-\mapsto hadrons$ [17,18].
These models can also provide qualitative,
and quantitative, information on the spectrum and masses of
mesonic and baryonic resonances [17,19].
Flux tube models have been employed
to estimate meson evaporation rates from quark-gluon plasmas which
are thought to form in heavy-ion collisions
[20]
and to
estimate baryon number penetrabilities
at the phase boundary between a quark-gluon plasma
and a hadron phase in a putative cosmic
first order QCD-phase transition [15].

In the Lund-model [18] the process
of $e^+e^-\mapsto hadrons$ is seen
as an annihilation of the $e^+e^-$-pair
into a quark-antiquark-pair ($q\bar q$).
A chromoelectric flux tube is formed between the
$q\bar q$-pair.
The maximum separation between the quark and the antiquark or,
equivalently, the maximum flux tube length $(x_m)$
depends on the
initial energy of the $e^+e^-$-pair ($E$). Both quantities are related
by
$$E=\sigma x_{m}\ ,\eqno(2)$$
where $\sigma\approx 0.177$ GeV$^2$
is the string tension of the flux tube.
Hadrons are then formed by successive $q\bar q$-pair creations
within the
chromoelectric field of the flux tube.

This process has its analogue in QED. It is known that there is
a finite probability for pair creation of an $e^+e^-$-pair
in a strong external electric field [21].
The electron and positron will arrange
themselves in such a way that the electric field is partially
screened and weakened.
The rest mass of this \lq\lq real\rq\rq\
$e^+e^-$-pair can be provided by the decrease in
the external electric field energy.

In QCD a chromoelectric
field between a $q\bar q$-pair of colors ($1\bar 1$)
can be screened in two different ways:
{\bf (a)} by the pair creation of a $q\bar q$-pair with the same color
as the original $q\bar q$-pair; or {\bf (b)} by two successive pair creations
of $q\bar q$-pairs with colors ($2\bar 2$) and ($3\bar 3$).
In general, processes {\bf (a)} and {\bf (b)} will happen multiple
times within a long flux tube.
Pair creation will stop
when the energy in the chromoelectric field
is exhausted. At this point, all quarks and antiquarks are arranged in
such a way that only residual chromoelectric fields remain.
This arrangement corresponds to a configuration of several color-singlet
hadrons. Thus the energy of the chromoelectric field has been converted
to rest mass and kinetic energy of hadrons in the final state.
Note that process {\bf (a)} corresponds to the formation of two mesons
and process {\bf (b)} corresponds to the formation of
a baryon-antibaryon pair.

Consider now the phase boundary between a quark-gluon plasma
and a hadron phase in a cosmic first order QCD-phase transition.
We can identify three microscopic mechanisms for the transport of
baryon number from
the quark-gluon plasma over the phase boundary into the hadron phase.
{\bf (I)} An energetic quark of color ($1$) passes over the phase boundary
into the hadron phase. The quark's color is screened
by the nearest plasma antiquark
of color ($\bar 1$). A flux tube forms between this
quark-antiquark pair and
the flux tube \lq\lq decays\rq\rq\
via two successive pair creations of quarks
with colors ($2\bar 2$) and ($3\bar 3$).
A color-singlet baryon
(consistent of three quarks with colors $123$)
forms in the hadron phase and three antiquarks
(colors $\bar 1\bar 2\bar 3$) pass back into the quark-gluon phase.
{\bf (II)} An energetic quark of color ($1$)
passes over the phase boundary into the hadron
phase and its color is screened by two close by
plasma quarks of colors ($2$) and ($3$).
The leading quark of color ($1$) forces (drags) the two screening quarks
into the hadron phase. A baryon is formed in the hadron phase.
{\bf (III)} A pre-formed \lq\lq baryonic cluster\rq\rq\ of three quarks
which initially resides in the quark-gluon plasma
passes over the phase boundary as a unit into the hadron phase.
The energy of this cluster exceeds the threshold energy for the
formation of a baryon in the hadron phase,
and the result is again the formation of a baryon.
To some extent mechanisms {\bf (II)} and {\bf (III)}
can be thought of describing
the same process. The efficiency of these processes is hard to
estimate since we are uncertain about the pre-formation probability
of \lq\lq baryonic clusters\rq\rq\ within the quark-gluon plasma.

In the following we estimate the efficiency of mechanism {\bf (I)} with
the help of a chromoelectric flux tube model.
Since we neglect baryon number transport by mechanisms {\bf (II)}
and {\bf (III)} we can only
obtain a lower limit for the baryon number penetrability,
$\langle\Sigma_q\rangle$. We can relate this lower limit on
$\langle\Sigma_q\rangle$ to a lower limit on
$\langle\Sigma_h\rangle$ if we use detailed balance in equation (1)
and take $\kappa =1$.

We can imagine a quark of color ($1$) which passes over
an idealized discontinuous phase boundary into the hadron phase
at time $t=0$. As already described above, a flux tube will
form between
this leading quark and a plasma antiquark of color ($\bar 1$).
The flux tube will continuously increase its length and slow the
motion of the leading quark until all the kinetic energy of the
quark is converted into energy of the flux tube. At this time, $t_0$,
the leading quark will reverse its motion and move back in the
direction of the phase boundary.

There is, however, a
probability that the flux tube decays by $q\bar q$-pair creation
before time $t_0$. We denote the probability that the flux tube
decays before time $t$ by $P_d(t)$. This probability satisfies
$$dP_d(t)=k_bV(t)\bigl(1-P_d(t)\bigr)dt+k_mV(t)
\bigl(1-P_d(t)\bigr)dt\ ,\eqno(3)$$
where $V(t)$ denotes the volume of the flux tube at time $t$ and
$k_m$ denotes the probability per unit time and unit volume for the
decay of the flux tube by a single $q\bar q$-pair creation.
Similarly, $k_b$ denotes the probability for the decay of the flux
tube by two successive $q\bar q$-pair creations.
In our calculation we will only consider the decay of flux tubes
by either a single pair creation of a $q\bar q$-pair with colors
($1\bar 1$) or two successive pair creations of $q\bar q$-pairs with
colors ($2\bar 2$) and ($3\bar 3$). We will not include additional
pair creations. This approximation is reasonable for the environment
under consideration, since only a very small fraction of thermal
quarks within the quark-gluon plasma will have enough energy to
create flux tubes in which multiple $q\bar q$-pair creations are
possible.
The probability, $P_b(t)$, that the flux tube decays via two successive
$q\bar q$-pair creations before time $t$ satisfies
$$dP_b(t)=k_bV(t)\bigl(1-P_d(t)\bigr)dt\ .\eqno(4)$$
Equations (3) and (4) can be solved to yield:
$$P_b(t)={k_b\over {k_b+k_m}}\Bigl(1-\exp\bigl(-(k_b+k_m)\int
\limits_0^tV(t')dt'\bigr)\Bigr)\ .\eqno(5)$$

We expect the formation of a baryon in the hadron phase
when three conditions are met: {\bf
(a)}
the flux tube decays via two successive $q\bar q$-pair creations;
{\bf (b)}
the energy of the leading quark exceeds the energy
threshold for baryon formation;
and {\bf (c)} the flux tube decays
while the leading quark still has momentum which is directed
away from the phase boundary, i.e. before time $t_0$.
The three antiquarks of colors ($\bar 1\bar 2\bar 3$) will
pass back into the quark-gluon plasma since there is
not enough energy to produce the rest masses of both a baryon and an
antibaryon in the hadron phase.

With the help of this simple prescription we can
obtain the baryon formation probability or, equivalently,
the baryon number penetrability, if we solve
for the dynamics of the quarks during this process.
This probability should be given by $P_b(t_0)$ from equation (5)
whenever the energy of the leading quark exceeds the threshold
for baryon formation.
We can approximate the length of the flux tube to be given at any time
by the shortest distance between the leading quark and the phase
boundary. That is, we assume that the flux tube is perpendicular
to the phase boundary.

We denote the perpendicular distance of the leading quark from
the phase boundary as $x_{\bot}$. Additionally, we denote the lateral position
coordinate {\it along} a phase boundary as $x_{\|}$.

Energy conservation ($E$) and momentum conservation
for the momentum component parallel to the phase boundary ($p_{\|}$) yield
$$E=const={{\sigma x_{\bot}}\over {(1-\dot x_{\|}^2)^{1/2}}}+
{m_q\over {(1-\dot x^2)^{1/2}}}\ ,\eqno(6a)$$
$$p_{\|}=const={{(\sigma x_{\bot})\dot x_{\|}}\over
{(1-\dot x_{\|}^2)^{1/2}}}+{{m_q\dot x_{\|}}\over {(1-\dot x^2)^{1/2}}}\ .
\eqno(6b)$$
In these expressions $x^2=x_{\|}^2+x_{\bot}^2\ $, while a dot over any quantity
indicates a time derivative, and $m_q$ is the quark current mass.
These equations imply that the quark velocity
parallel to the phase boundary ($\dot x_{\|}$) is conserved.
We can solve the equation of motion and compute the integral
$\int\limits_0^{t_0}V(t)dt=\pi{\Lambda}^2\int\limits_0^{t_0}
x_{\bot}(t)dt$ , where $\Lambda$ is the radius of the flux tube.
In the limit of small quark masses $(m_q/E)\mapsto 0$ we obtain
$$\int\limits_0^{t_0}V(t)dt={\pi\Lambda^2\over 2\sigma^2}
E^2\cos \theta\ ,\eqno(7)$$
with $\theta$ the angle of incidence of the leading
quark relative to the phase boundary.
Using equations (5) and (7) we can determine
the baryon formation probability,
$P_b(E,\theta )\equiv P_b(t_0)$, for a quark of energy
$E$ which approaches the phase boundary at an incident
angle $\theta$.

In order to form a baryon in the hadron phase the energy
of the leading quark has to exceed a threshold
energy.
The leading quark has to have enough energy to produce
the baryon rest mass of ($m_b \sim 1 \ GeV$) and the kinetic
energy for the motion of the baryon parallel to the phase boundary.
The second contribution to this energy threshold
follows from the
conservation of velocity parallel to the phase boundary.
The energy threshold condition is given by
$$E>E_{th}={m_b\over \cos\theta}\ .\eqno(8a)$$

It is conceivable that the energy threshold exceeds the one given
in equation (8a). This is because
the initial state includes a quark and an antiquark in the
quark-gluon
plasma (colors $1\bar 1$), whereas the final state
includes a baryon in the hadron phase and
three antiquarks (colors $\bar 1\bar 2\bar 3$)
in the quark-gluon plasma.
There is an associated average interaction energy for each
quark and antiquark which resides in the quark-gluon plasma.
We can estimate this average energy if we approximate the
quark-gluon plasma with the help of the bag models.
Such an approximation gives
$E_{int}\approx 3.7T$ [5] for the average energy.
The final state in the process of baryon formation includes an
additional quark.
It is not known to
what extent a cooling of the quark-gluon plasma could
provide the interaction energy for the additional quark.
We could modify the threshold condition in equation (8a) to be
$$E>E_{th}={m_b\over cos\theta}+Bn_q^{-1}\ .\eqno(8b)$$
The correct energy threshold should be between the values
given in equations (8a) and (8b).

We can now compute the thermal average
of the baryon number penetrability at the phase
boundary in a cosmic QCD-phase transition
if we assume that mechanism {\bf (I)} is the dominant mechanism
for baryon number transport across the phase boundary. We find
$${\langle\Sigma_h\rangle}={{1\over {f_{b-\bar b}}}\int\limits
_0^{\pi}d\theta\int\limits_{E_{th}}^{\infty}dE\,
{dn_{q-\bar q}\over
{dEd\theta}}\,{\dot x}_{\bot}^q(\theta)\,P_b(E,\theta)}
\ ,\eqno(9)$$
where $n_{q-\bar q}$ is the excess density in quarks, i.e.
the density of quarks minus the density of antiquarks, and
$f_{b-\bar b}$ is the excess flux in baryons as in equation (1).
The differential excess quark number density for quarks in a given energy
interval $(dE)$ and in a given interval of incident angles
$(d\theta)$ is
$${dn_{q-\bar q}\over {dEd\theta}}={{{\mu_q\over T}\,{g_q\over {2\pi^2}}}\,
{{E^2\exp {(E/T)}}\over {({\exp(E/T)}+1)^2}}\sin \theta}\
.\eqno(10)$$
In this expression the quantities $\mu_q$, $g_q$,
and $T$ are the quark chemical
potential, the statistical weight of quarks, and the temperature,
respectively.

Equation (10) assumes that an isotropic and massless
quark Fermi gas obtains in the quark-gluon plasma during the transition.
The statistical weight
of quarks, $g_q$, is the product of the number of
colors (3) times the number of relativistic quark flavors
(probably 2, the up and down quark) times
the number of possible quark spins (2). The excess flux in baryons
within the hadronic phase, $f_{b-\bar b}$, is given by
$${f_{b-\bar b}}={{\mu_b\over T}{g_b\over {2\pi^2}}\,{(m_b+T)}\,
{T^2}{\exp(-m_b/T)}}\ .\eqno(11)$$
In equation (11) $m_b$, $\mu_b$, and $g_b$ are the baryon mass, baryon
chemical potential, and baryon statistical weight, respectively.
Note that $g_b=4$ and $\mu_b=3\mu_q$.
The component of the quark velocity perpendicular to the
phase boundary is simply
$${\dot x}_{\bot}^q(\theta)=\cos\theta\ .\eqno(12)$$

In order to obtain a quantitative result
for $\langle\Sigma_h\rangle$ from equations (5-9) we need to estimate
the values for the quantities $a\equiv k_b/(k_m+k_b)$ and
$b\equiv\pi\Lambda^2(k_b+k_m)/2\sigma^2$.
In the limit, $k_b<<k_m$, these quantities can be approximated
by $a\approx k_b/k_m$ and $b\approx \pi\Lambda^2k_m/2\sigma^2$.

In principle the value for the quantity $b$ can be inferred
by an extension of the $e^+e^-$-pair creation probability
in an electric field in QED. This pair creation
probability is accurately known [21].
In the case of QCD the $q\bar q$-pair creation probability
within a flux tube is
$$b\approx {\pi\Lambda^2k_m\over 2\sigma^2}=f_1\Lambda^2\sum_{n=1}
^{\infty}{1\over n^2}exp\bigl(-f_2{m_q^2n\over \sigma}\bigr)\ ,
\eqno(13)$$
with $f_1$ and $f_2$ numerical factors and $m_q$ the
quark mass. The pair creation probability given by equation
(13) has been compared to results of
$e^+e^-\mapsto hadrons$ experiments. There is disagreement
in the literature about the numerical values of
$f_1$ and $f_2$. In addition, the values for the flux tube width,
$\Lambda$, and the quark mass, $m_q$, are not accurately known.
In the present analysis we will treat $b$ as a parameter.
We assume a value for $b$ between $b=0.32$ GeV$^{-2}$
[17] and
$b=0.05$ GeV$^{-2}$ [20].
In comparison Sumiyoshi {\it et al.} [15] used the rather small
value of $b=0.044$ GeV$^{-2}$.

We can infer the value for the quantity $a\approx k_b/k_m$
(i.e. for the ratio of the probability of flux tube decay
via two successive $q\bar q$-pair creations to the probability
of flux tube decay via a single $q\bar q$-pair creation) from
experiments of $e^+e^-\mapsto hadrons$.
We associate the production of a baryon-antibaryon pair with the
decay of a flux tube via two successive $q\bar q$-pair creations.
Similarly, we associate the production of two mesons
with the decay of a flux tube via a single $q\bar q$-pair creation.
A naive estimate of the quantity $a$ would then be the ratio
of the number of baryons to the number of mesons produced
in $e^+e^-$-annihilations. This ratio would be approximately
$a\approx 1/20$, a value consistent with the small ratio which has been used by
Sumiyoshi {\it et al.}. We will argue that the appropriate
value for $a$ is much larger than $a\approx 1/20$.
In fact, we think that a better estimate for this quantity is
$a\approx 1/5-1/3$. This ratio is also in agreement with a
suppression factor of $a\approx 1/5$ for the production of baryons
relative to the production of mesons which has been deduced
by Casher {\it et al.} [17].

There are two effects for an enhancement of meson production compared
to that for baryon production in accelerator $e^+e^-$-annihilations.
First, flux tubes of small length and energy do not provide enough
energy for the production of a baryon-antibaryon pair. These flux
tubes necessarily have to decay into mesons. Second, baryonic
resonances eventually decay into a baryon of lower rest mass.
In most cases this decay is accompanied by the production
of up to three mesons. Thus the total number of mesons and baryons
produced in $e^+e^-$-annihilations does not directly reveal the
ratio for the flux tube decay probabilities
$a\approx k_b/k_m$ which are relevant to our problem.

In $e^+e^-$-annihilations at center-of-mass energy $E_{CM}=29$
GeV [22] a shower of hadrons with
energies varying between $E=E_{CM}/2$ and $E=m_{\pi}$ is produced.
Baryons, of course, are only formed between energies of
$E=E_{CM}/2$ and $E=m_p$, with $m_p$ the proton mass.
For an approximate determination of the quantity $a$ it is
appropriate to compare the number of produced baryons to the
number of produced mesons at fixed baryon and meson energy,
$E\geq m_p$ . In this way low-energy mesons ($E\leq m_p$)
which are produced by the decay of short flux tubes or by the decay
of baryonic resonances are not counted.
If this is done, a ratio of $a\approx 1/3$ - $1/5$ is
found to obtain over a wide range of energies.

There is another effect which might further enhance
the relevant
value for $a$ in the cosmic QCD-phase transition. The thermal
average in equation (9) strongly favors energies slightly above
the energy threshold for baryon formation.
It is known that baryon production cross sections are normally larger
near the threshold energy than at higher energies
and, thus, the formation of baryons at the phase boundary might
be enhanced compared to that in accelerator $e^+e^-$-annihilations.
The existence
of a multitude of baryonic resonances in the energy range
$1$ GeV $<$ $E$ $<$ $2$ GeV could also imply an anomalously large
effective value for the quantity $a$ in the QCD-phase transition.

We have performed a numerical computation to obtain the thermal
average of the baryon number penetrability,
$\langle\Sigma_h\rangle$, given by equation (9).
Our results are displayed in Figures 1 and 2. Both figures show
the thermal average of the baryon number penetrability
times the ratio of probabilities $(k_m/k_b)$. In these figures we plot
$\langle\Sigma_h\rangle (k_m/k_b)$
as a function of QCD-phase transition
temperature $T$.
The thermal average of the baryon number penetrability,
$\langle\Sigma_h\rangle$, is then easily obtained if we multiply
the results shown in Figures 1 and 2 by the ratio $(k_b/k_m)$.
As discussed above, this ratio is approximately
$a\approx (k_b/k_m)\approx 1/5-1/3$.
In both figures we show results for different values of the
parameter $b\approx \pi\Lambda^2k_m/2\sigma^2$, ranging between
$b=0.4$ GeV$^{-2}$ and $b=0.05$ GeV$^{-2}$.
In Figure 1 we have assumed the threshold condition of equation (8a), while
in Figure 2 we have assumed the threshold condition of equation (8b).

Our results imply that the baryon number penetrability $\langle\Sigma_h\rangle$
may be smaller than unity.
The baryon number penetrability is found to range between
$\langle\Sigma_h\rangle\sim 0.01$ and
$\langle\Sigma_h\rangle\sim 0.1$, depending on the quantities
$a$ and $b$, the threshold condition, and the phase transition
temperature. We do not find, however, baryon number penetrabilities
as small as $\langle\Sigma_h\rangle\sim 10^{-3}$.
Such small values for $\langle\Sigma_h\rangle$ have been
suggested by Sumiyoshi {\it et al.} [15] and Fuller {\it et al.} [5].
Since our results neglect several mechanisms to form
baryons at the phase boundary (i.e. mechanisms
{\bf (II)} and {\bf (III)} from
above), and since we find mechanism {\bf (I)} to be quite efficient
we conclude that values for $\langle\Sigma_h\rangle$ as small as
$\langle\Sigma_h\rangle\sim 10^{-3}$ are rather unlikely.

However, we think it is essential to point out inherent uncertainties
involved in any flux tube calculation of the baryon number
penetrability
at the phase boundary between a quark-gluon plasma
and a hadron phase. We could argue that such calculations are
necessarily oversimplified and that the results are model dependent.
Several critical uncertainties and assumptions in
these calculations are as follows:
the assumption of an idealized discontinuous phase boundary;
the assumption of a noninteracting nuclear density environment
in the phases;
the oversimplified classical treatment of the dynamics of leading and
screening quarks; the neglect of other mechanisms to
transport baryon number across the phase boundary; and the
uncertainties in parameters which describe the flux tube and
its decay.

Any serious calculation of the baryon number penetrability at
the phase boundary should take into account the finite
extension and physical nature of the phase boundary itself.
Unfortunately, not much is known about the phase boundary.
The width of the phase boundary should be roughly of the
order of a Debye color screening length, $L_D\sim 1$fermi.
In comparison the extension of flux tubes formed by
energetic quarks ($E\sim 1$ GeV) is of the same order,
$L_f\sim 1$fermi.
Thus hadronization will most likely occur {\it within} the phase
boundary.

The environment under consideration is at nuclear
densities in both phases. This environment is compared to
results obtained in accelerators ($e^+e^-\mapsto hadrons$)
which effectively operate in \lq\lq vacuum\rq\rq .
It is conceivable that hadronization at the phase boundary is a
\lq\lq collective\rq\rq process which involves a large
number of particles.

A serious calculation of $\langle\Sigma_h\rangle$ should also
include well known quantum mechanical principles.
In our calculation of the dynamics of the quark
position we implicitly assumed a simultaneous knowledge of
quark position and momentum.
Thus, we treated the motion of the quarks completely classically.
This, of course, is not correct
for microscopic systems such as two quarks separated by
roughly, $L_f\sim 1$fermi. A better estimate of baryon number penetrability
would be a solution to the Dirac equation for the motion of a quark
near a plane parallel, discontinuous potential distribution. This potential
would be
located at the phase boundary and,
in a more sophisticated treatment, could be taken as linearly
increasing in the direction towards the hadron phase.
We would have to allow for the decay
of the potential, which would simulate the decay of the flux tube.
Such a calculation would take better account of the quantum
mechanical nature of the problem.

The third and fourth critical points in our list have already
been briefly discussed above. We suggested alternative mechanisms for
baryon number transport across the phase boundary. We are not
able to give estimates for the efficiency of these mechanisms.
This is mainly due to a poor knowledge of baryon pre-formation
probabilities
within the quark-gluon plasma.

Possible ranges for
the phenomenological quantities $a$ and $b$ have been given. The
exact values which should enter into our calculations are uncertain.
We have extracted the value for $a$ and $b$ from
$e^+e^-\mapsto hadrons$ experiments, i.e. from environments
which are substantially different from the environment in a
cosmic QCD-phase transition. There is some hope that results of
heavy-ion collision experiments could remove some of these
uncertainties. However, even this environment is significantly
different from the environment encountered in the early universe.
This is because heavy-ion collisions are performed at large net
baryon number densities and under conditions which are far from
thermodynamic equilibrium, in contrast to a first order QCD-phase
transition in the early universe.
In view of these uncertainties we believe that flux tube
models can only be used to estimate the order of magnitude
of the baryon number penetrability at the phase
boundary in a cosmic QCD-phase transition.

\vskip 0.15in
\centerline{\bf 3. Baryon-Baryon Coalescence}
\vskip 0.1in

In this section we will briefly discuss analogies of the problem of
baryon number transport across the phase boundary in a
cosmic QCD-phase transition to issues in nuclear physics.
We will attempt to estimate the baryon number penetrability,
$\langle\Sigma_h\rangle$, by employing
{\bf (a)} nonrelativistic nucleon-nucleon potentials and
{\bf (b)} results of a possible explanation for an unresolved phenomena
in deep inelastic scattering experiments.

A
phenomenological understanding of the
internal structure of baryons can be obtained from bag models
(for a review see [23]). In bag models baryons
and baryonic resonances are described by a
Fermi-\lq\lq sea\rq\rq\ of three quasi-free quarks existing in a
perturbed vacuum. This perturbed vacuum is of finite spatial extension
and is referred to as the \lq\lq bag\rq\rq .
The size of the bag is of
the order of one Fermi, the approximate dimension of a
baryon. The energy or rest mass of the baryon is then given by the
sum of rest masses and kinetic energies of three quarks, the energy
of the perturbed vacuum, and the surface energy associated with the
surface between perturbed vacuum (the bag) and
\lq\lq ordinary\rq\rq vacuum.
A macroscopic quark-gluon plasma phase could be approximated
by an extension of the MIT-bag model [5].
It would consist of a large number of quarks ($N$) existing in a
macroscopic region of perturbed vacuum, i.e. in a large bag.

Consider the probability $\langle\Sigma_h\rangle$
for a baryon which approaches the phase
boundary (from the hadronic phase) to dissociate into its
constituents (three quarks) and to pass across the phase boundary
into the quark-gluon plasma. This process can be approximated
as the coalescence of a three-quark bag (the baryon)
with an $N$-quark bag (the quark-gluon plasma).

This view suggests the following question. What is the probability
for two baryons to form a \lq\lq di-baryon\rq\rq ,
i.e for two three-quark bags to coalesce and form a
six-quark bag. The answer to this question could be intimately
related to the probability, $\langle\Sigma_h\rangle$.

Before we try to exploit the suggested analogy we want to
establish an essential difference between the coalescence of two
baryons
and the coalescence of a baryon
with a macroscopic quark-gluon phase. This difference is
given by the geometries of the two processes.
There are two effects pertaining to the simple understanding of baryons from
the MIT-bag
model. First, the decrease in bag surface area
in the process of baryon-baryon coalescence is smaller
than the decrease in surface area in the process of coalescence
of a baryon with a quark-gluon phase. The associated decreases
in surface
energy are thus different for the two processes and we expect a higher
probability for the coalescence of a baryon with a quark-gluon
plasma than for the coalescence of two baryons.
Second, in the formation of a six-quark bag
the Pauli-exclusion principle plays an important role.
Whenever the two baryons in the initial state include two quarks
with the same
flavor, color, and spin (the probability for this to happen is
roughly 1/2) then
one of these quarks would have to occupy an
excited state within the final state six-quark bag.
An excited state in the six-quark bag,
in general, will lie several hundred MeV higher in energy than
the ground state. For energetic reasons, this state might not then
be accessible to the quark, and the formation of a six-quark
bag would be Pauli blocked.
In the case of a macroscopic
quark-gluon plasma in the early universe we expect a quasi-continuum in energy
states,
so that Pauli blocking is never complete.
For these reasons we expect the probability for the coalescence
of two baryons
to be smaller than the probability
for the coalescence of a baryon with a quark-gluon plasma.

\vskip 0.15in
\noindent
{\bf 3a. Nuclear Core Potentials}
\vskip 0.1in

It would be interesting if there was an inhibition in the
formation of di-baryons (or six-quark bags) observed in nuclear
physics experiments.
Such an inhibition could imply an anomalously low value for
$\langle\Sigma_h\rangle$.
It is common nuclear physics knowledge that the saturation of
nuclear bulk matter can only be explained by the existence of a
strong repulsion between nucleons at short separation distances
(cf. [24]). To this end a \lq\lq hard core\rq\rq ,
which is an infinitely high potential barrier at a nucleon-nucleon
separation distance of
$r_c\approx 0.5$F is introduced into two-body
nucleon-nucleon potentials. There is also an indication for the
repulsion of two nucleons at short separation distances from
nucleon-nucleon
scattering data and deuteron properties [25].

Unfortunately, there are a great number of phenomenological,
nonrelativistic nucleon-nucleon potentials which can explain
nucleon-nucleon scattering data and deuteron properties.
This multitude of parameter-fitted nucleon-nucleon potentials reflects our
inability to deduce the correct nucleon-nucleon potential from
first principles.
Most of these phenomenological potentials employ a nuclear
\lq\lq hard core\rq\rq . However, in a few cases potentials
have been modified to incorporate a {\sl finite} height potential
barrier at short distances which is referred to as a nuclear
\lq\lq soft core\rq\rq .
By how much could the potential barrier between
two nucleons at short separation distances
be reduced before a conflict with nucleon-nucleon scattering data
is inevitable?

Bressel, Kerman, \& Rouben [26] modified the Hamada-Johnston
nucleon-nucleon potential by replacing the nuclear
\lq\lq hard core\rq\rq with a \lq\lq soft core\rq\rq .
They extended the core region to $r_c\approx 0.7$F and introduced
a finite potential barrier of height $V_0\approx 600$ MeV at $r_c$.
Lacombe {\it et al.} [27] changed the Paris nucleon-nucleon potential
by incorporating a finite potential barrier of height
$V_0\approx 200$MeV
which extends to a radius of $r_c\approx 0.8$F. In these potentials
the \lq\lq soft core\rq\rq\ potential barrier at $r_c$ is, however,
spin and isospin
dependent. Both potentials are in good agreement
with nucleon-nucleon scattering data
and deuteron properties.

There are two relevant quantities in nuclear physics which have been
well determined experimentally:
the root-mean-square radius of the deuteron and the nucleon-nucleon
scattering
length. Here the nucleon-nucleon scattering length is the square root of the
zero energy nucleon-nucleon scattering cross section.
It seems that any potential which includes a nuclear
\lq\lq hard core\rq\rq is unable to explain both quantities
simultaneously
[28]. Most of these potentials overestimate the root-mean-square
radius of the deuteron or, if they explain the root-mean-square radius
correctly, they underestimate the scattering length. An obvious cure
should be the introduction of a nuclear \lq\lq soft core\rq\rq .
A nuclear \lq\lq soft core\rq\rq would slightly lower the
root-mean-square radius of the deuteron. However, it would not affect
the nucleon-nucleon cross section at zero energy, i.e. the scattering
length.
Thus knowledge of the root-mean-square radius
of the deuteron and the nucleon-nucleon scattering length might
be valuable in obtaining additional information about the
nucleon-nucleon potential at short separation distances.

We would like to obtain a numerical estimate for the penetration
probability of two baryons, assuming a potential barrier of height,
$V_0$, at a baryon-baryon separation distance of $r_c$.
We will denote this probability by $\Sigma$. We can then seek a thermal
average of this quantity $\langle\Sigma\rangle$ which is
appropriate for a thermal distribution of baryons in the early
universe at the QCD-phase transition temperature.
As discussed above, this probability should be related to the
probability for the coalescence of a baryon with a quark-gluon
plasma, $\langle\Sigma_h\rangle$. To compute $\Sigma$
we have performed a simple quantum mechanical
computation to determine the tunneling probability
of a nonrelativistic baryon through a barrier of height, $V_0$, and
extension, $r_c$. In such a tunneling process the six-quark bag
would be the intermediate state.
Our results obtained in this manner
are displayed in Figures 3 and 4. In
these figures we show $\langle\Sigma\rangle$ as a function of the QCD-phase
transition temperature, $T$. We have employed different potential barriers
$V_0$ (ranging between 200 MeV and 800 MeV) in our estimates of
$\langle\Sigma\rangle$.

Figure 3 assumes a
potential barrier width of $d=2r_c=1$fermi; whereas Figure 4
assumes $d=1.4$fermi.
It is evident that in order to obtain a baryon-baryon penetration
probability as small as $\langle\Sigma\rangle=0.1$, the potential barrier
must be high ($V_0$ $^>_\sim$ 600 MeV) and the temperature must be low.
A potential barrier as low as $V_0\approx 200$ MeV [27]
results in a penetration probability of
$\langle\Sigma\rangle\sim 0.5$.
In view of the increased probability for the coalescence of a baryon
with a quark-gluon plasma compared to the coalescence of two baryons,
our results could be consistent with $\langle\Sigma_h\rangle =1$.

\vfill
\eject
\noindent
{\bf 3b. The EMC-Effect and Six-Quark Bags}
\vskip 0.1in

In deep inelastic lepton-nucleus scattering experiments
a nontrivial deviation
of the cross section for scattering of leptons off big nuclei
(e.g. iron) compared to that for scattering of leptons
off small nuclei (e.g. deuterium) has been observed. This deviation can not be
attributed to differences in the Fermi motions of nucleons
(Fermi-gas model)
within different nuclei. The effect is known as the EMC-effect
[29] and the reader is referred to [30] for a detailed discussion.

It has been suggested that the EMC-effect could be explained
by a significant six-quark bag component in nuclei
[31,32,33].
Detailed estimates indicate
that a six-quark bag component of 16\% in $^3$He-nuclei could
explain the scattering data of leptons scattering off $^3$He
[32]. Similarly, a six-quark bag component of 30\% in
$^{56}$Fe-nuclei could account for the data observed in leptons
scattering off $^{56}$Fe [33].
Note, however, that a potential six-quark bag component in nuclei
is not the only explanation for the EMC-effect.

If we assume that a six-quark bag component in $^3$He is the sole explanation
of the EMC-effect, we can obtain a simple estimate for the
baryon-baryon penetration probability in nuclei.
To this end, we imagine three nucleons confined to the volume of
a $^3$He-nucleus as frequently colliding. The fraction of the time during
which we would observe a six-quark bag in $^3$He, $P_6$,
is then given by the
total nucleon-nucleon collision frequency, $P_c$,
times the lifetime of a
six-quark bag, $\tau_6$, times the probability for penetration of two
nucleons, $\Sigma$. We can then write,
$$P_6\approx P_c\tau_6\Sigma\ .\eqno(14)$$

An estimate of the nucleon-nucleon collision frequency
can be obtained by assuming a nucleon of cross sectional
area $\pi a^2$ moving through a nuclear volume $(4\pi /3)r^3$ at
a typical velocity $v$.
The total nucleon-nucleon collision frequency
is then simply three times
the reciprocal of the time which a single nucleon
needs to
traverse the whole nuclear volume. The factor of 3 enters the
calculation since there exist three distinct
ways to form a pair
of nucleons in $^3$He.
We can approximate the radius of a $^3$He-nucleus to be $r\approx 2$fermi,
and the radius of a nucleon to be $a\approx 0.5$fermi.
An approximate nucleon velocity
can be obtained either from the
Heisenberg uncertainty principle, or from the momentum expectation value
of a nucleon moving in the mean field of a $^3$He-nucleus.

If we estimate the nucleon-nucleon collision frequency, $P_c$,
in this manner we can
rewrite equation (14) as
$$\Sigma\approx {2\over 5}{mr^4\over\hbar a^2}{P_6\over\tau_6}\ ,\eqno(15)$$
where $m$ is the nucleon mass and $\hbar$ is Planck's constant.
For lack of a more sophisticated treatment, we assume that the
lifetime of a six-quark bag is
$\tau_6\approx 3\times 10^{-24}$s,
a typical strong interaction time.
This is also the time light needs to travel over the
dimensions of a nucleon.
Finally, if we
take $P_6\approx 0.16$ [32], we can infer a
nucleon-nucleon
penetration probability $\Sigma$.
With this set of assumptions we obtain
the surprising result that $\Sigma\approx 20$!
This value can not be correct, since $\Sigma$ can not exceed unity.
Similarly, we obtain a value for $\Sigma$ which is above unity
if we perform an analogous
calculation for the nucleus of $^{56}$Fe [33].

Obviously, our assumptions and approximations in this crude estimate must break
down.
We can imagine two
simple ways to resolve the issue. It could be that
the existence of a six-quark bag
component in nuclei is {\sl not} the sole explanation
of the EMC-effect. The actual
value for
$P_6$ would then be much smaller than the value
given by Pirner \& Vary [32] and
Carlson \& Havens [33]. It is also conceivable that the life time
of six-quark bags in nuclei,
$\tau_6$, is longer than our estimate
$\tau_6\approx 3\times 10^{-24}$s. If, however, both of these
assumptions are correct, i.e. the six-quark bag component
in nuclei is substantial and the life time of such six-quark bags
is short, then we have to seek the explanation for the large
value of $\Sigma$ in our crude approximations. In this case
our results for $\Sigma$ would be consistent with a baryon-baryon
penetration probability of order unity.

\vfill
\eject
\centerline{\bf 4. Conclusions}
\vskip 0.15in

We have estimated the efficiency of baryon number transport across the
phase boundary
between a quark-gluon plasma phase and a hadron phase in a putative
cosmic first order QCD-phase transition. Such a transition is likely
to form
baryon number inhomogeneities. The amplitude of these baryon number
inhomogeneities could reach very high values if there is a substantial
inhibition for the transport of baryon number across the phase
boundary.

In our calculations we have employed phenomenological flux tube models
and analogies of the problem of baryon number transport across
the phase boundary to issues in nuclear physics.
We have discussed appreciable uncertainties in computations
of the baryon number penetrability. We have argued that such calculations
should only be used to indicate the order of magnitude of baryon
number penetrabilities. Our results indicate baryon number
penetrabilities at the phase boundary which could be consistent
with a maximum possible probability of unity.

\vskip 0.15in
\centerline{\bf 5. Acknowledgements}
\vskip 0.1in
The authors are grateful for useful discussions with R. J. Gould,
D. Kaplan,
A. Manohar, S. Klarsfeld, H. Paar, H. J. Pirner, and R. Swanson.
This work was supported in part by NSF Grant PHY91-21623 and by an
IGPP minigrant. It was also performed in part under the auspices of the
U.S. Department of Energy by the Lawrence Livermore National Laboratory
under contract number W-7405-ENG-48 and DoE Nuclear Theory Grant SF-ENG-48.
\vfill
\eject

\centerline{\bf References}
\vskip 0.1in
\item{\bf [1]}
E. Witten, Phys. Rev. D 30 (1984) 272.
\item{\bf [2]}
J.H. Applegate and C.T. Hogan, Phys. Rev. D 31 (1985) 3037.
\item{\bf [3]}
 J.H. Applegate, C.T. Hogan, and R.J. Scherrer, Phys. Rev. D 35
(1987) 1151; J.H. Applegate, C.T. Hogan, and R.J. Scherrer, Astrophys. J. 329
(1988) 592.
\item{\bf [4]}
C.R. Alcock, G.M. Fuller, and G.J. Mathews, Astrophys. J. 320 (1987) 439.
\item{\bf [5]}
G.M. Fuller, G.J. Mathews, and C.R. Alcock, Phys. Rev. D 37 (1988)
1380.
\item{\bf [6]}
H. Kurki-Suonio, R.A. Matzner, J. Centrella, T. Rothman, and J.R. Wilson,
Phys. Rev. D 38 (1988) 1091; R.A. Malaney and W.A. Fowler, Astrophys. J. 333
(1988) 14; H. Kurki-Suonio and R.A. Matzner, Phys. Rev. D 39 (1989) 1046;
N. Terasawa and K. Sato, Prog. Theor. Phys. 81 (1989) 254; N. Terasawa and
K. Sato, Phys. Rev. D 39 (1989) 2893; T. Kajino and R.N. Boyd,
Astrophys. J. 359 (1990) 267; G.J. Mathews, B.S. Meyer, C.R. Alcock, and
G.M. Fuller, Astrophys. J. 358 (1990) 36; K. Jedamzik, G.M. Fuller, and
G.J. Mathews, Astrophys. J. 423 (1994) 50.
\item{\bf [7]}
C.R. Alcock and E. Farhi, Phys. Rev. D  32 (1985) 1273;
C.R. Alcock and A. Olinto, Ann. Rev. Nuc. Part. Science 38
(1988) 161.
\item{\bf [8]}
R. Schaeffer, P. Delbougo-Salvador, and J. Audouze, Nature 317
(1985) 407; J. Madsen and K. Riisager, Phys. Lett. B 158 (1985) 208;
K. Jedamzik, G.M. Fuller, G.J. Mathews, and T. Kajino, Astrophys. J. 422
(1994) 423.
\item{\bf [9]}
L.J. Hall and S.D.H. Hsu, Phys. Rev. Lett. 64 (1990) 2848.
\item{\bf [10]}
J.M. Quashnack, A. Loeb, and D.N. Spergel, Astrophys. J. Lett. 344 (1989) L49.
\item{\bf [11]}
F.R. Brown {\it et al.}, Phys. Rev. Lett. 65 (1990) 2491.
\item{\bf [12]}
B. Petersson, Nucl. Phys. A 525 (1991) 237c.
\item{\bf [13]}
S.A. Bonometto and O. Pantano, Phys. Rep. 228 (1993) 175.
\item{\bf [14]}
K. Freese and F. Adams, Phys. Rev. D 41 (1990) 2449;
B. Link, Phys. Rev. Lett. 68 (1992) 2425;
P. Huet, K. Kajantie, R.G. Leigh, B.H. Liu, and L. McLerran, Phys. Rev. D
48 (1993) 2477.
\item{\bf [15]}
K. Sumiyoshi, T. Kusaka, T. Kamio, and T. Kajino, Phys. Lett. B 225 (1989) 10.
\item{\bf [16]}
J. Kogut and L. Susskind, Phys. Rev. D 11 (1975) 395.
\item{\bf [17]}
A. Casher, H. Neuberger, and S. Nussinov, Phys. Rev. D 20 (1979) 179.
\item{\bf [18]}
B. Anderson, G. Gustafson, and T. Sjostrand, Phys. Rep. 97 (1983) 32.
\item{\bf [19]}
A.B. Migdal, S.B. Khokhlachev, V.Y. Borne, Phys. Lett. B 228 (1989) 167.
\item{\bf [20]}
B. Banerjee, N.K. Glendenning, and T. Matsui, Phys. Lett. B 127 (1983) 453.
\item{\bf [21]}
J. Schwinger, Phys. Rev. 82 (1951) 664.
\item{\bf [22]}
Aihara {\it et al.}, Phys. Rev. Lett. 11 (1988) 1263.
\item{\bf [23]}
A.W. Thomas, in {\it Advances in Nuclear Physics}, edited by J.W. Negele
and E. Vogt (Plenum, New York) 13 (1983) 1.
\item{\bf [24]}
A.L. Fetter and J.D. Walecka, {\it Quantum Theory of Many-Particle Systems},
McGraw-Hill 1971.
\item{\bf [25]}
R. Jastrow, Phys. Rev. 81 (1951) 165.
\item{\bf [26]}
C.N. Bressel, A.K. Kerman, and B. Rouben, Nucl. Phys. A 124 (1968) 624.
\item{\bf [27]}
M. Lacombe {\it et al.}, Phys. Rev. D 12 (1975) 1495.
\item{\bf [28]}
S. Klarsfeld {\it et al.}, Nucl. Phys. A 456 (1986) 373.
\item{\bf [29]}
J.J. Aubert {\it et al.}, Phys. Lett. B 123 (1983) 275.
\item{\bf [30]}
E.L. Berger and F. Coestner, Ann. Rev. Nucl. Part. Science 37 (1987) 463.
\item{\bf [31]}
E. Lehman, Phys. Lett. B 62 (1976) 296; H. Faissner, B.R. Kim, and
H. Reithler, Phys. Rev. D 30 (1984) 900.
\item{\bf [32]}
H.J. Pirner and J.P. Vary, Phys. Rev. Lett. 46 (1980) 1376.
\item{\bf [33]}
C.E. Carlson and T.J. Havens, Phys. Rev. Lett. 51 (1983) 261.

\vfil\eject

\centerline{\bf Figure Captions}
\vskip 0.1in
\item{\bf Figure 1} The thermal average of the baryon number
penetrability $\langle\Sigma_h\rangle$ times the probability ratio
$(k_m/k_b)$ as a function of temperature $T$ in MeV. This product is
shown for four different values of the parameter $b$, ranging
between $b=0.05$ GeV$^{-2}$ and $b=0.4$ GeV$^{-2}$. In this figure
we have assumed the energy threshold condition given in equation
(8a).

\item{\bf Figure 2} Same as Figure 1 but here we have used the energy
threshold condition given in equation (8b).

\item{\bf Figure 3} The thermal average of the baryon-baryon penetration
probability $\langle\Sigma\rangle$ as a function of temperature
$T$ in MeV. Results are shown for square-wave potential barriers with
barrier heights $V_0=$ 200 MeV, 400 MeV, 600 MeV, and 800 MeV.
The spatial width of the square-wave barrier is assumed to be
$d=1$fermi.

\item{\bf Figure 4} Same as Figure 3 but here we have assumed a width of
the square-wave potential barrier of $d=1.4$fermi.

\vfil
\eject
\end